\documentclass[a4paper, biblatex]{jacow}

\usepackage{graphicx}
\usepackage{amsmath}
\usepackage{siunitx}
\usepackage{makecell}
\usepackage{soul,xcolor}

\begin{document}

\title{Compensating for Amplifier Non-Linearity in a SEL Controller
\thanks{The authors of this work grant the arXiv.org and LLRF Workshop's International Organizing Committee a non-exclusive and irrevocable license to distribute the article, and certify that they have the right to grant this license.}\thanks{ Work supported by FermiForward Discovery Group, LLC under Contract No. 89243024CSC000002 with U.S. DOE.}}

\author{%
  S. Raman\thanks{ssankar@fnal.gov}, P. Varghese, L. Reyes, M. Guran \\
  Fermi National Accelerator Laboratory (FNAL), Batavia, IL 60510, USA \\[6pt]
  L. Doolittle (Retd.), S. Murthy, Q. Du \\
  Lawrence Berkeley National Laboratory (LBNL), Berkeley, CA 94720, USA
}

\maketitle

\begin{abstract}
The Self-Excited Loop (SEL) architecture, used in some continuous-wave (CW) superconducting linacs, relies on a positive feedback mechanism that requires carefully defined operating limits to ensure stable operation. These limits are typically derived from amplifier calibration, which characterizes the relationship between forward power and DAC drive. However, amplifier non-linearity often prevents a simple linear fit of this characteristic, introducing errors that can compromise stability. To address this, we present a modified calibration procedure that incorporates amplifier non-linearity into the SEL framework. The approach is validated with test data from a 32 kW solid-state amplifier (SSA) and a cavity emulator developed for the Fermilab PIP-II linac. 
\end{abstract}

\section{INTRODUCTION}
Self-Excited Loop (SEL) controllers are used in continuous-wave (CW) superconducting linac to regulate the cavity field. The SEL relies on positive feedback, which means the loop has to be carefully calibrated to stay stable while still responding quickly to changes such as cavity detuning and microphonics. The calibration process ensures that the amplifiers deliver the precise range of amplfier power needed for beam stability and optimal performance. For this reason, the accuracy of the amplifier calibration plays a central role in the SEL architecturre.

In an ideal scenario, the relationship between forward power and the cavity DAC drive is linear, allowing straightforward calibration and predictable system behavior. However, in practical systems, some solid-state amplifiers (SSAs) exhibit non-linear behavior, deviating from the ideal linear response. This non-linearity requires an accurate characterization so that the drive level for any desired field can be correctly computed.

In this paper, we present a modified calibration procedure that compensates for amplifier non-linearity within the SEL controller framework. The approach is validated using test data from a \SI{32}{\kilo\watt} solid-state amplifier developed for the Fermilab PIP-II superconducting linac. By incorporating the amplifier non-linearity into the calibration process, our method improves the accuracy of forward power mapping enabling stable operation of the SEL controller.

\section{CALIBRATION OF SSA}
\subsection{Self-Excited Loop Architecture}
\begin{figure}[h]
  \centering
  \includegraphics[width=0.7\linewidth]{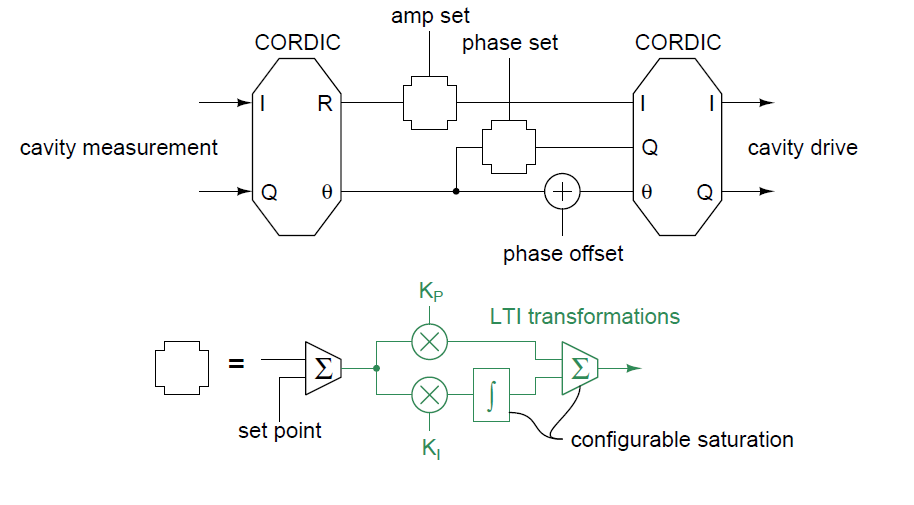}
  \caption{SEL architecture.}
  \label{fig:sel_arch}
\end{figure}

The self-excited loop (SEL) architecture represents a digital implementation of the classical analog oscillator scheme introduced in Delayen’s early work[1]. The feedback core of this architecture, shown in Fig.~\ref{fig:sel_arch}, begins with a CORDIC block that converts the measured in-phase (I) and quadrature (Q) signals into magnitude and phase, which feed into independent feedback loops for amplitude and phase.

Unlike conventional feedback paths, where the outputs are directly transformed back to I and Q, the SEL architecture employs a rotation process that is central to its operation. Specifically, the amplitude and phase controller outputs are reinterpreted as new I and Q components, but rotated by the sum of the measured cavity phase and a phase offset. This offset provides the positive feedback condition that enables self-excited loop oscillation of the cavity drive. At resonance, the phase feedback setpoint stays close to zero, corresponding to particle acceleration near the crest of the cavity field.

The I and Q inputs to the second CORDIC produce the forward power in terms of its real and imaginary components. The operating limits for these components can be configured for the different SEL operating modes. Determining these limits requires a series of power and amplifier calibrations, which are described in the following sections.

\subsection{Gradient Calibration}
The calibration process begins with a gradient calibration, which establishes a mapping between the requested cavity gradient and measurable RF quantities. This step provides the scaling needed to express the accelerating voltage, stored energy, and forward power in consistent physical units, forming the basis for the subsequent SSA calibration.

The signal calibration starts with the requested accelerating gradient, $E$, which is converted into the effective accelerating voltage,
\begin{equation}
  V = E \times l \; \si{\mega\volt},
\end{equation}
where $E$ is the cavity gradient in MV/m and $l$ is the cavity length in metres.

From this definition, the stored energy can be calculated as
\begin{equation}
  \sqrt{U} = \frac{V}{\sqrt{\tfrac{r}{Q} \, 2 \pi f_0}} \, \si{\sqrt{\joule}}.
\end{equation}

The stored energy is then related to the required forward power as
\begin{equation}
  \sqrt{P} = \sqrt{U} \times \sqrt{\frac{\pi f_{0}}{2 Q_{L}}} \, \si{\sqrt{\watt}} ,
\end{equation}
where $f_{0}$ is the cavity frequency and $Q_{L}$is the loaded quality factor.

Finally, the target forward power ADC-normalized amplitude, is given by
\begin{equation}
  Fwd_{t} = \frac{\sqrt{P}}{fwd_{\text{fs}}}.
\end{equation}

\renewcommand{\arraystretch}{1}
\begin{table*}[t]
  \caption{Saturation Limits for SEL Modes}
  \label{tab:sat_limits}
  \centering
  \begin{tabular}{|c|c|c|c|c|}
    \hline
    Mode & $X_{lo}$ & $X_{hi}$ & $Y_{lo}$ & $Y_{hi}$ \\
    \hline
    SEL Raw & -- & -- & -- & -- \\  \hline
    SEL & $SSA_{Slope} \times Fwd_{t}$ & $SSA_{Slope} \times F_{trgt}$ & 0 & 0 \\  \hline
    SELA & $SSA_{Slope} \times Fwd_{t} \times 0.85$ & $(SSA_{Slope} \times Fwd_{t} + ped) \times 1.15$ & 0 & 0 \\  \hline
    SELAP & $SSA_{Slope} \times Fwd_{t} \times 0.85$ & $(SSA_{Slope} \times Fwd_{t} + ped) \times 1.15$ & $-policy_Y$ & $policy_Y$ \\
    \hline
  \end{tabular}
\end{table*}

\subsection{SSA Calibrations}
The SSA is calibrated by driving it with a staircase sequence of DAC levels and recording the forward I and Q signals using the ADCs and the downconverter. The magnitude and phase are computed from these signals and are used to characterize the SSA

The calibration then reduces to estimating a median slope relating DAC drive levels to measured forward amplitudes. If the measured forward signal (represented as normalized ADC values from Eq. 4) is denoted as $x$ and the DAC drive as $y$, this slope is defined as
\begin{equation}
  SSA_{Slope} = \operatorname{median}\left(\frac{y}{x}\right).
\end{equation}

The corresponding calibration target is then
\begin{equation}
  DAC_{Drive} = SSA_{Slope} \times Fwd_{t}.
\end{equation}

\subsection{SEL Modes and Operating Limits}
The SEL architecture supports four modes of operation: SEL Raw, SEL, SELA, and SELAP. The SELAP configuration, corresponds to a self-excited loop with both amplitude and phase feedback loops closed. This mode replicates the behavior of the generator-driven resonator (GDR) mode, when the cavity detuning is sufficiently small to permit locking. However, unlike conventional GDR, the control system reverts to SELA when detuning exceeds the locking range. SELA represents a self-excited loop with amplitude regulation while tracking the cavity’s natural resonant frequency. In contrast, SEL is an open-loop configuration in which the system simply follows the cavity resonance without amplitude or phase stabilization. Finally, SEL Raw operates similarly to SEL, but the drive amplitude is specified as a percentage of the full scale drive rather than calibrated values in MV. This mode is primarily intended for early operation, before full signal calibration is available.

\begin{figure}[h]
  \centering
  \includegraphics[width=0.7\linewidth]{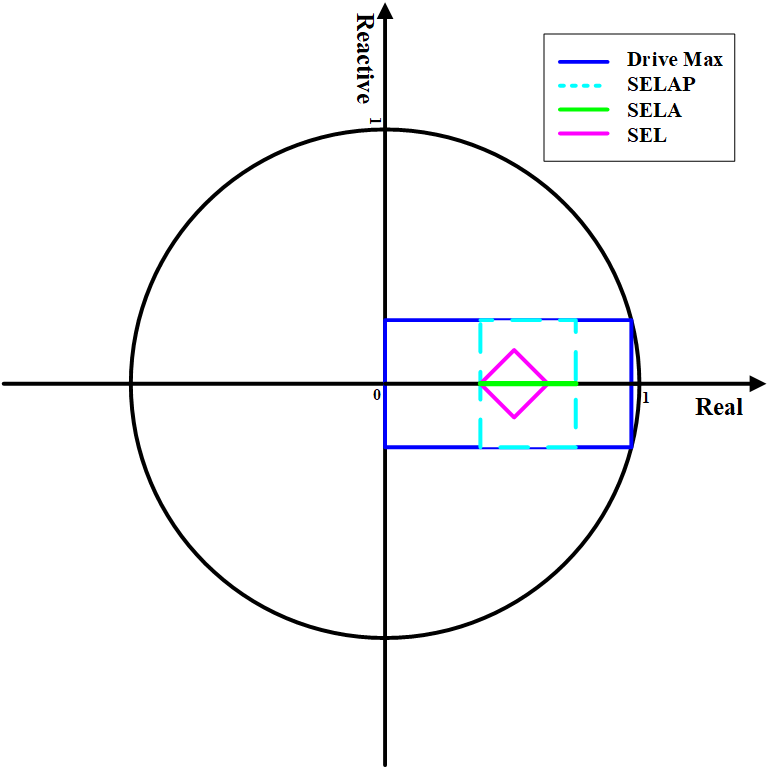}
  \caption{Drive limits.}
  \label{fig:drive_limits}
\end{figure}

Having established the functional distinctions among the SEL modes, their practical operating limits are defined through the calibration process. These limits are summarized in Table \ref{tab:sat_limits}, where each mode applies specific bounds to the I/Q drive coordinates, and SELAP imposes additional phase loop constraints. This can be pictorially represented as shown in Fig.~\ref{fig:drive_limits}.

\begin{figure}[h]
  \centering
  \includegraphics[width=0.7\linewidth]{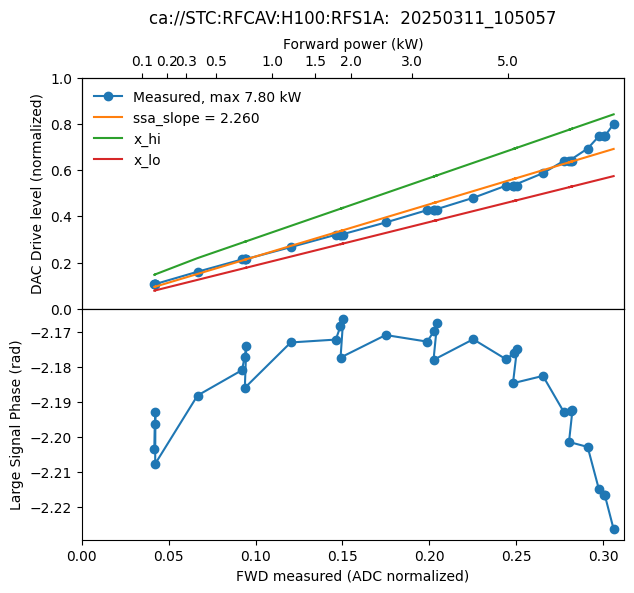}
  \caption{SSA calibration.}
  \label{fig:ssa_cal}
\end{figure}

In this framework, $X_{hi}$ and $X_{lo}$ denote the amplitude boundaries for the SSA drive, as summarized in Table \ref{tab:sat_limits} and illustrated in Fig.~\ref{fig:ssa_cal}. These limits are computed directly from the calibrated SSA slope, $SSA_{Slope}$ (from Eq. 5), and the target ADC-normalized amplitude (Eq. 4). The lower amplitude limit is defined as
\begin{equation}
  X_{lo} = 0.85 \times SSA_{Slope} \times Fwd_{t},
\end{equation}
while the upper limit incorporates a pedestal offset, $ped$ to account for baseline output or low-level leakage,
\begin{equation}
  X_{hi} = 1.15 \times (SSA_{Slope} \times Fwd_{t} + ped).
\end{equation}

For small amplitudes, a low-slope approximation is used to refine the upper limit,
\begin{equation}
  low_{slope} = \frac{SSA_{Slope} \cdot x_{min} + ped}{x_{min}},
\end{equation}
\begin{equation}
  X_{hi} = \min\bigl(X_{hi},  low_{slope} \times Fwd_{t} \times 1.15\bigr),
\end{equation}
where $x_{min}$ is the minimum normalized forward amplitude used for low-power slope calculations. This ensures that the operating range remains within approximately $\pm 15\%$ of the calibrated linear response.

Similarly, the phase loop saturation limits, $Y_{hi}$ and $Y_{lo}$, are derived from the DAC-normalized drive components. Using the maximum drive magnitude, $Drv_{max}$, and the reactive component, $Drv_{imag}$, the effective phase constraints are computed as:
\begin{equation}
  policy_{Y} = Drv_{\max} \times {\sqrt{Drv_{imag}}},
\end{equation}
\begin{equation}
  policy_{X} = Drv_{\max} \times {\sqrt{1-Drv_{imag}}}.
\end{equation}

Here, $policy_{X}$ represents the maximum allowable real component of the drive across all operating modes. The phase feedback loop is engaged exclusively in SELAP mode, where the phase setpoint is defined by the user as the target phase. These limits define the allowable range of phase and amplitude for the SSA, and any measured points outside these boundaries are highlighted during calibration to validate linearity and safe operation of the amplifier.

A successful calibration is demonstrated when the SSA’s DAC drive levels produce a forward response that falls within the expected linear limits. Plots of drive versus forward power typically show a straight-line region, providing both validation of the calibration and an intuitive view of amplifier behavior across its usable range. While this framework assumes linear operation, measurements performed on the PIP-II HB650 test stand SSAs reveal noticeable deviations from ideal behavior. These observations highlight the importance of examining the non-linear response characteristics of the amplifiers, which are discussed in the following section.

\section{SSA NON-LINEARITY}

\renewcommand{\arraystretch}{1}
\begin{table*}[t]
  \caption{SSA Operating Limits for SEL Modes Using Affine Calibration}
  \label{tab:ssa_limits_new}
  \centering
  \begin{tabular}{|c|c|c|c|c|}
    \hline
    Mode & $X_{lo}$ & $X_{hi}$ & $Y_{lo}$ & $Y_{hi}$ \\
    \hline
    SEL Raw & -- & -- & -- & -- \\    \hline
    SEL & $SSA_{Slope} \cdot Fwd_{t}$ & $SSA_{Slope} \cdot Fwd_{t}$ & 0 & 0 \\    \hline
    SELA & \makecell[l]{$0.85 \cdot SSA_{Slope} \cdot Fwd_{t}$\\
                        $+ SSA_{Offset}$} &
           \makecell[l]{$\min\Big((SSA_{Slope} \cdot Fwd_{t} + ped) \cdot 1.15$\\
                        $+ SSA_{Offset},\; X_{hi2},\; Drv_{max}\Big)$} &
           0 & 0 \\    \hline
    SELAP & \makecell[l]{$0.85 \cdot SSA_{Slope} \cdot Fwd_{t}$\\
                         $+ SSA_{Offset}$} &
            \makecell[l]{$\min\Big((SSA_{Slope} \cdot Fwd_{t} + ped) \cdot 1.15$\\
                         $+ SSA_{Offset},\; X_{hi2},\; Drv_{max}\Big)$} &
            $-policy_Y$ & $policy_Y$ \\
    \hline
  \end{tabular}
\end{table*}

\begin{figure}[h]
  \centering
  \includegraphics[width=0.7\linewidth]{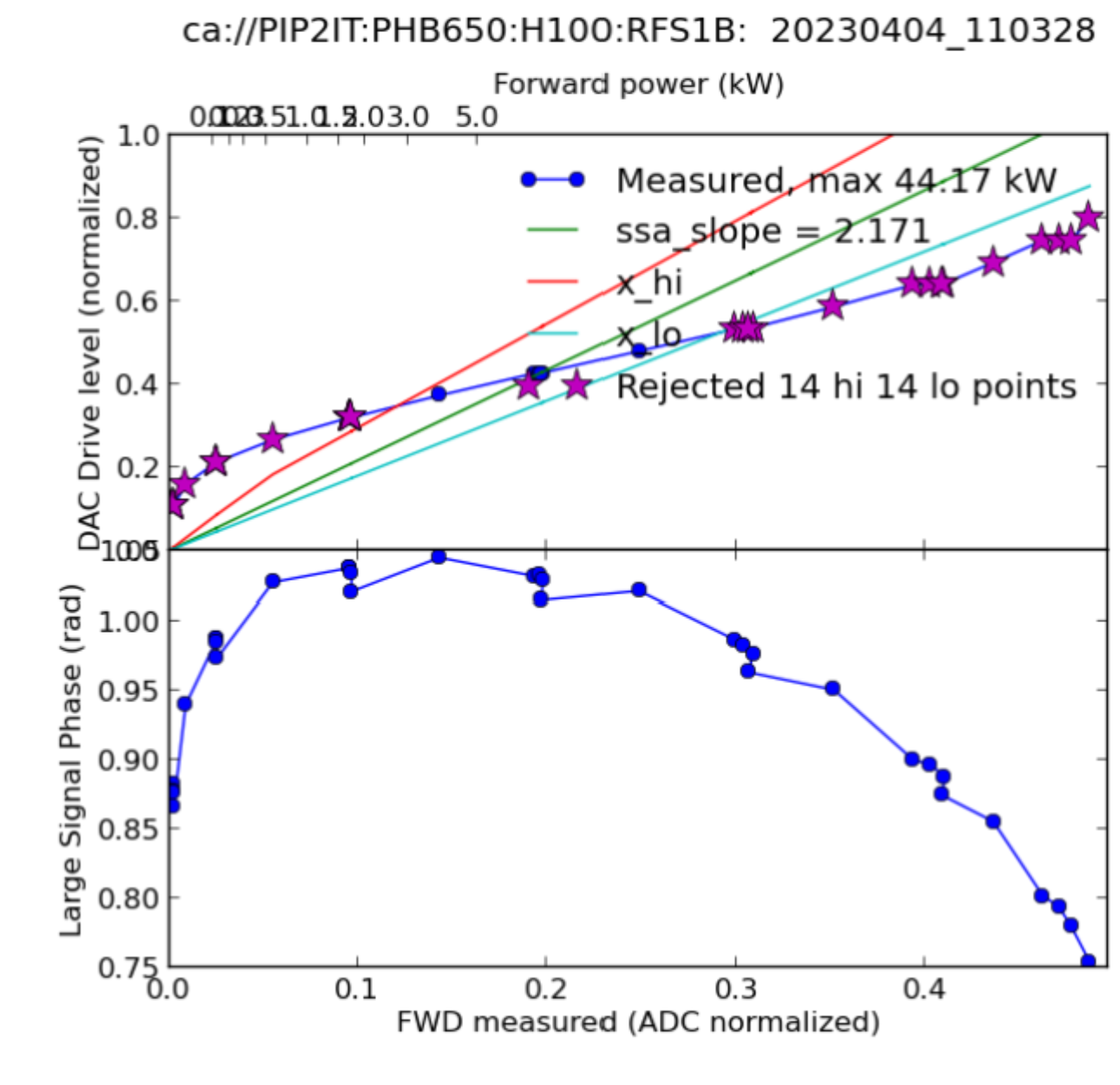}
  \caption{Observed SSA non-linearity.}
  \label{fig:ssa_cal_nonlin_linfit}
\end{figure}

The calibration approach described above assumes that the amplifier behaves linearly, with a proportional relationship between DAC drive levels and measured forward power that can be well-approximated by a linear fit passing through the origin. This assumption holds in many cases and provides a practical means to establish operating limits for SEL operation. However, experience with SSA in the PIP-II HB650 cryomodule test stand has shown that this linear approximation does not apply to the nonlinear amplifier characteristic shown in Figure~\ref{fig:ssa_cal_nonlin_linfit}.

The HB650 cryomodule test stand measurements of the \SI{32}{\kilo\watt} SSAs revealed clear departures from linear calibration at relatively low drive levels. In these cases, the measured forward power no longer followed the expected straight-line relationship, and a number of calibration points were rejected because they fell outside the allowable tolerance, causing the calibration to fail. Figure~\ref{fig:ssa_cal_nonlin_linfit} shows representative results, where the amplifier’s response bends away from the linear fit well before reaching the upper end of the drive range. This indicates that the simple slope-based calibration is insufficient to represent the full behavior of these amplifiers.

When the amplifier response deviates from the assumed linear relationship, the controller may apply an incorrect drive level, that may result in improper operation. Accurately capturing these deviations through higher-order models is therefore essential to maintain precise gradient control and ensure reliable, repeatable operation of the cryomodule.

\section{COMPENSATING FOR THE NON-LINEARITY}
\subsection{Affine Calibration Scheme}
The original calibration method treated the SSA response as a purely proportional relationship between the DAC drive and the measured forward power. In other words, it assumed a straight line through the origin with slope only. In practice, measurements showed that the SSA characteristic response curve is shifted away from the origin, creating a vertical offset that cannot be described by slope alone. To correct for this effect, the calibration algorithm was updated with two main changes: the use of a SSA minimum power, $P_{\min}$ in watts and the use of a linear fit that includes both slope and offset.

The new parameter $P_{\min}$ sets the minimum forward power that is considered when fitting the calibration line. Points below this threshold are ignored, since the SSA response at low power is more affected by non-linear distortions and measurement noise. This ensures that the fit is based only on the part of the curve where the SSA behaves more predictably.

Once the dataset is filtered, the calibration no longer fits a slope-only line. Instead, it uses an affine model,
\begin{equation}
  y = SSA_{Slope} \times x + SSA_{Offset},
\end{equation}
where $x$ is the measured forward signal and $y$ is the DAC drive command. Here, $SSA_{Slope}$ reflects how much the forward signal grows with increasing DAC drive, while $SSA_{Offset}$ corrects for the baseline output that remains when the drive is zero.

\begin{figure}[h]
  \centering
  \includegraphics[width=0.7\linewidth]{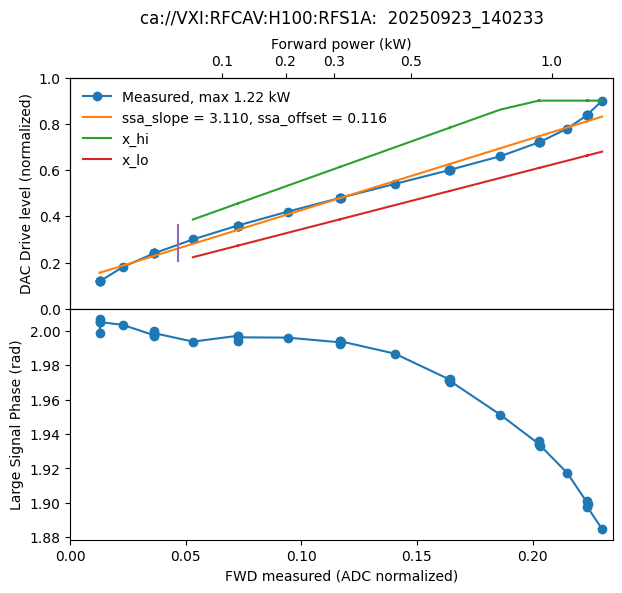}
  \caption{Non-linear fit of SSA.}
  \label{fig:ssa_cal_nonlin}
\end{figure}

These two parameters directly influence the construction of the acceptance limits. The upper and lower bounds are no longer simple $\pm 15\%$ deviations around a zero-intercept line, but instead scaled envelopes around the affine model. The offset shifts the entire calibration line vertically, while the slope determines the envelope width as a function of $x$. This is shown in Fig.~\ref{fig:ssa_cal_nonlin}.

In practice, this yields limits of the form
\begin{equation}
\begin{split}
  X_{lo} &= 0.85 \times (SSA_{Slope} \times Fwd_{t}) + SSA_{Offset} \\
  &\quad, \\
  X_{hi} &= 1.15 \times (SSA_{Slope} \times Fwd_{t} + ped) + SSA_{Offset}.
\end{split}
\end{equation}

The upper limit \(X_{hi}\) is further constrained by a lowslope correction, defined as
\begin{equation}
  X_{hi2} = low_{slope} \times Fwd_{t} \times 1.15,
\end{equation}
which ensures that at low amplitudes, the DAC drive remains within a safe operating range. In addition, $X_{hi}$ is limited by the maximum allowable DAC drive, $Drv_{\max}$, to prevent exceeding the amplifier’s hardware capability. The final upper limit is therefore the minimum of these three values:
\begin{equation}
  X_{hi} = \min(X_{hi}, X_{hi2}, Drv_{\max}).
\end{equation}

These adjustments allow the SSA calibration to accurately capture non-linear behavior across a drive range, providing a safe and reliable operating envelope for SEL operation. The resulting operating limits for the different SEL modes are summarized in Table \ref{tab:ssa_limits_new}.

\subsection{Prototype Piecewise Calibration Scheme}
As an alternative to the earlier calibration approach, we propose a piecewise linear model to better accommodate non-linear amplifier behavior. In this formulation, the independent variable $x$ denotes the normalized measured forward ADC value (Eq. 4), and the dependent variable $y$ denotes the normalized DAC drive level. The model partitions the operating range into a low-power region and a high-power region. The transition occurs at $x = x_{min}$.

In the low-power region ($x < {x_{min}}$), the output drive is given by
\begin{equation}
  y =  low_{slope} \times x,
\end{equation}
and the corresponding tolerance limits are
\begin{equation}
\begin{split}
  X_{lo}^{low} &= 0.85 \times (low_{slope} \times Fwd_{t}), \\
  X_{hi}^{low} &= 1.15 \times (low_{slope} \times Fwd_{t}).
\end{split}
\end{equation}

For $x \ge x_{min}$, the model switches to the high-power segment,
\begin{equation}
  y = high_{slope} \times x + offset,
\end{equation}
with the tolerance bounds expressed as
\begin{equation}
\begin{split}
  X_{lo}^{high} &= 0.85 \times (high_{slope} \times Fwd_{t}) + offset, \\
  X_{hi}^{high} &= 1.15 \times (high_{slope} \times Fwd_{t}) + offset.
\end{split}
\end{equation}

The two slopes are related by
\begin{equation}
   low_{slope} = \frac{high_{slope} \times x_{min} + offset}{x_{min}}.
\end{equation}

The offset term in the high-power formula shifts the baseline drive level to account for residual drive requirements or saturation effects in the high region.

\begin{figure}[h]
  \centering
  \includegraphics[width=0.7\linewidth]{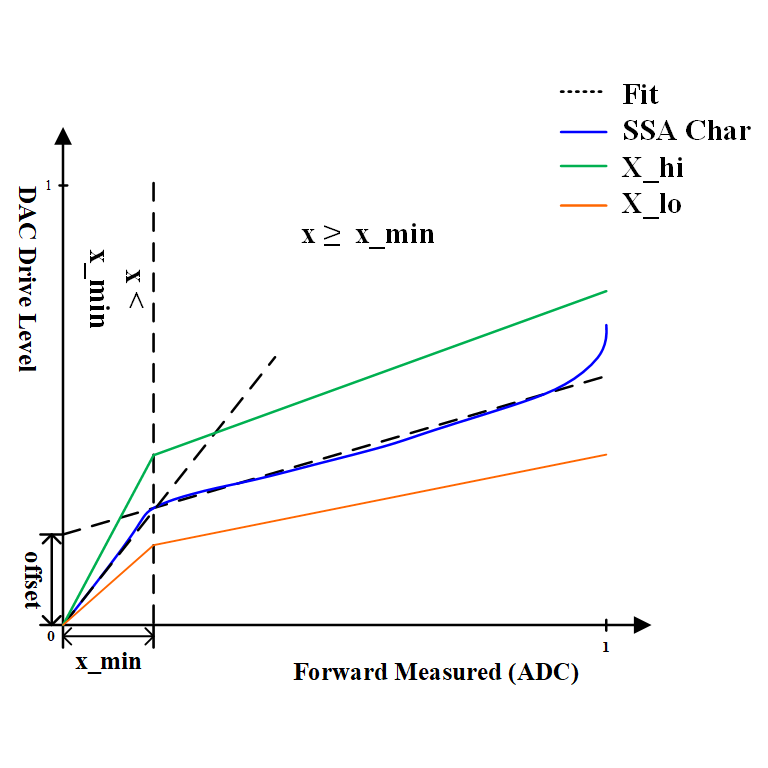}
  \caption{Graphical representation of the proposed piecewise method.}
  \label{fig:piecewise}
\end{figure}

To illustrate the proposed piecewise method, Fig.~\ref{fig:piecewise} presents a pictorial representation of its operational principle. The graph highlights the segmentation of the input-output relationship and the corresponding slopes applied in each region, providing a clear visualization of how the method approximates the system response. While this approach is still at the prototype stage, the figure demonstrates the key idea and its potential for implementation in future tests.

This piecewise construction affords flexibility: the slope and offset in each region adaptively reflect amplifier behavior, while the limit bounds ${X_{lo}^{low}, X_{hi}^{low}, X_{lo}^{high}, X_{hi}^{high}}$ scale naturally with those local operating slopes. At present, this approach remains a prototype and has not yet been implemented. However, it offers a improved method for resolving the nonlinearities. By explicitly distinguishing between $low_{slope}$ and $high_{slope}$ behavior and embedding offset into the model, the piecewise fit captures the SSA response with greater accuracy and robustness, ultimately improving the reliability of cavity operation under both low and high power conditions.

\section{CONCLUSION}
During operation of the PIP-II HB650 cryomodule test stand, a pronounced non-linearity was observed in the solid-state amplifier response, which exposed limitations in the existing calibration methodology. The previous approach, relying solely on a simple slope passing through the origin to relate DAC drive to measured forward amplitude, failed to capture the full non-linear behavior of the amplifier. As a result, the operational limits derived from this method were inaccurate, and the cavity could not be driven in SELA and SELAP modes, rendering it effectively inoperable.

To address this issue, a new calibration framework was developed based on an affine model that incorporates both the SSA slope and an SSA offset. The slope captures the rate at which the amplifier output grows with DAC drive, while the offset accounts for the baseline output, enabling a more precise representation of the amplifier’s true response. Using these two parameters, amplitude and phase limits can be calculated that accurately reflect the non-linear behavior of the SSA, providing a reliable operational envelope for the cavity.

In addition to this affine model, an alternate piecewise prototype has been proposed to handle amplifier nonlinearities by explicitly separating the low and high power regions. This method defines distinct slopes for the two operating ranges and uses an offset to capture residual drive effects in the high power region. Future work will evaluate its robustness, test its accuracy in hardware, and compare its performance against the affine calibration framework.


\begin{thebibliography}{5}   
\bibitem{}J. R. Delayen, “Phase and Amplitude Stabilization of Superconducting Resonators,” Ph.D. Thesis, Caltech (1978).
\bibitem{}J. Delayen, T. Allison, C. Hovater, J. Musson, and T. Plawski,"Development of a Digital Self-Excited Loop for Field Control in High-Q Superconducting Cavities”, in Proc. PAC’07, Albuquerque, NM, USA, Jun. 2007.
\bibitem{}L. Doolittle, S. Murthy, Q. Du, PIP-II LLRF Collaboration Documentation Discussions, 2024-2025.
\bibitem{}L. Doolittle, S. Murthy "Data flow for LCLS-II LLRF cavity amplitude setting", LBL  LLRF Note, August 2025.

\end{thebibliography}
\end{document}